\documentclass[useAMS,usenatbib]{mn2e}
\usepackage{times}
\usepackage{epsfig}

\def\DLO{{D_{\rm LO}}}
\def\DSL{{D_{\rm SL}}}
\def\DSO{{D_{\rm SO}}}
\def\crit{{\rm crit}}
\def\max{{\rm max}}

\title[Lensing by Black Holes in Galactic Nuclei]
{A Simple Model for Lensing by Black Holes in Galactic Nuclei}

\author[M.C. Werner and N.W. Evans] {M.C. Werner\thanks{E-mail:
mcw36@ast.cam.ac.uk; nwe@ast.cam.ac.uk} and
N.W. Evans\footnotemark[1]\\ Institute of Astronomy, University of
Cambridge, Madingley Road, Cambridge, CB3 0HA, United Kingdom}

\begin{document}

\date{This draft 2006 February 13}

\pagerange{\pageref{firstpage}--\pageref{lastpage}} \pubyear{0000}
\maketitle
\label{firstpage}

\begin{abstract}
The lensing properties of the Plummer model with a central point mass
and external shear are derived, including the image multiplicities,
critical curves and caustics. This provides a simple model for a
flattened galaxy with a central supermassive black hole.  For the
Plummer model with black hole, the maximum number of images is 4,
provided the black hole mass is less than an upper bound which is
calculated analytically. This introduces a method to constrain black
hole masses by counting images, thus applicable at cosmological
distance. With shear, the maximum number of images is 6 and we
illustrate the occurrence of an astroid caustic and two metamorphoses.
\end{abstract}



\section{Introduction}

Albert Einstein, in his 1911 paper \textit{On the Influence of
Gravitation on the Propagation of Light}, applied the principle of
equivalence to the deflection of light rays by the sun. This marks the
beginning of our modern understanding of gravitational lensing. After
the full prediction from General Relativity had been confirmed by
Eddington's observations of the 1919 solar eclipse, it was only in
1936 that Einstein considered, rather pessimistically, the deflection
due to stars. Zwicky's (1937) insight one year later that galaxies
were much better candidates was finally corroborated by the discovery
of the first doubly-lensed quasar in 1979 by Walsh, Carswell and
Weymann.

Nowadays, over a 100 multiply imaged gravitational lens systems are known. In many of
these, a background quasar or radio galaxy is lensed into doublets or
quadruplets by an early-type galaxy.  Provided the density
distribution at the centre of the lensing galaxy is weakly singular
(that is, no worse than $\rho \sim 1/r$), then gravitational lensing
yields an odd number of images, generally three or five. If the
singularity is stronger than this, then there is an even number of
images, as the central image is absent \citep{Ev98}. Most early-type
galaxies are believed to harbour supermassive black holes in their
centres. The existence and the magnification of the central image in
gravitational lens systems are very sensitive to the mass distribution
in the centre and the presence of a black hole. \cite{Ke03} has shown
that the likely range of the flux of any central image is below the
range of current instruments, but within the grasp of forthcoming, high
sensitivity VLBI arrays. Thus, it is becoming possible to probe the
mass distribution in the central tens of parsecs of the lensing
galaxy. It is therefore worthwhile to develop some simple models of
lensing by galaxies with black holes. \cite{Ma01} already studied a
galaxy model akin to the cored isothermal sphere and showed that a
central black hole -- provided it is not too massive -- can create an
additional faint central image.

The present paper is a mainly analytical study of the Plummer lens
with a central black hole and in the presence of external shear. In \S
2, we briefly review the properties of the spherical Plummer lens and
show that its images obey a magnification invariant. This model is
extended in \S 3 to include a central point mass to which external
shear is added in \S 4.

\begin{figure*}
\centering
\includegraphics[width=0.65\columnwidth, height=0.65\columnwidth]{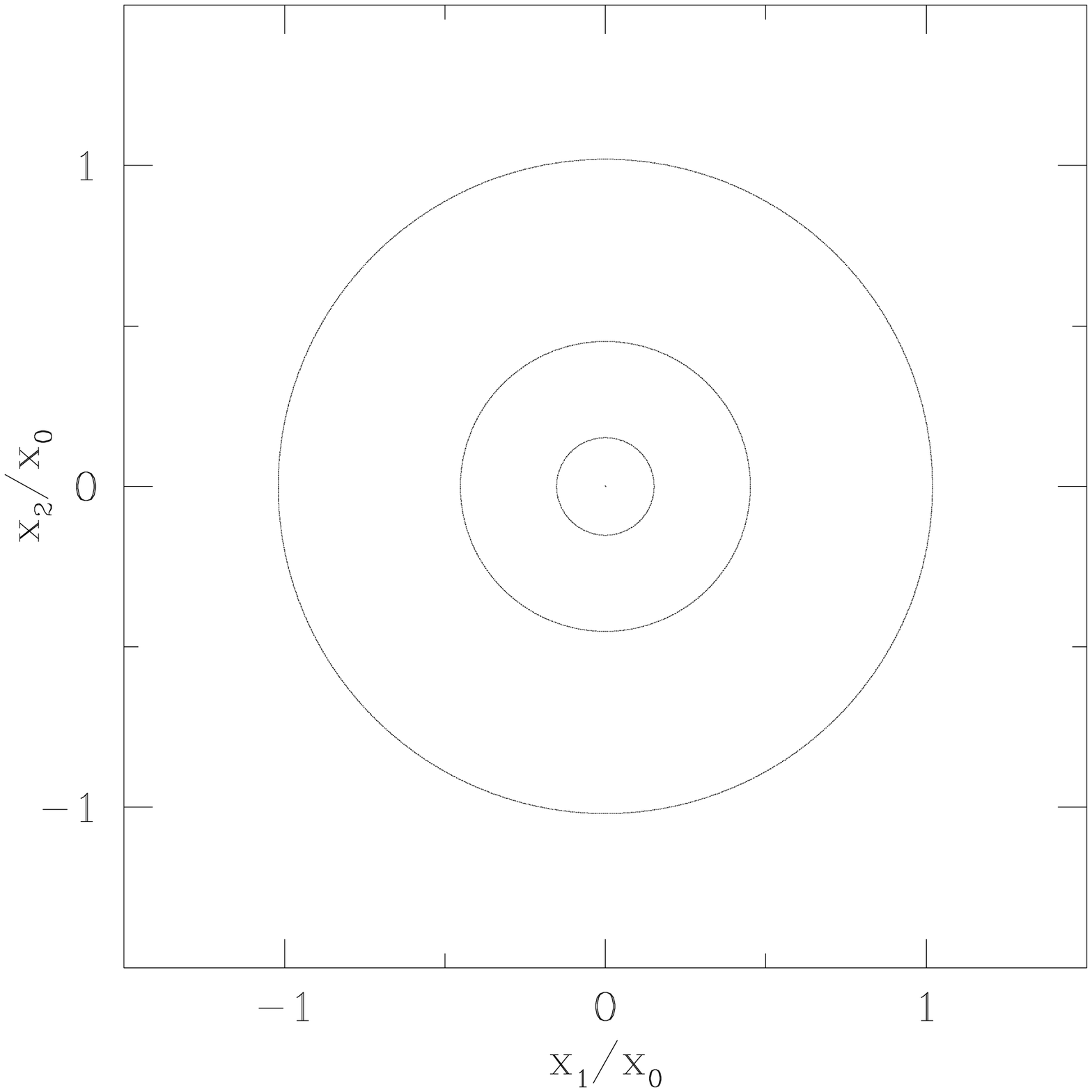}
\includegraphics[width=0.65\columnwidth, height=0.65\columnwidth]{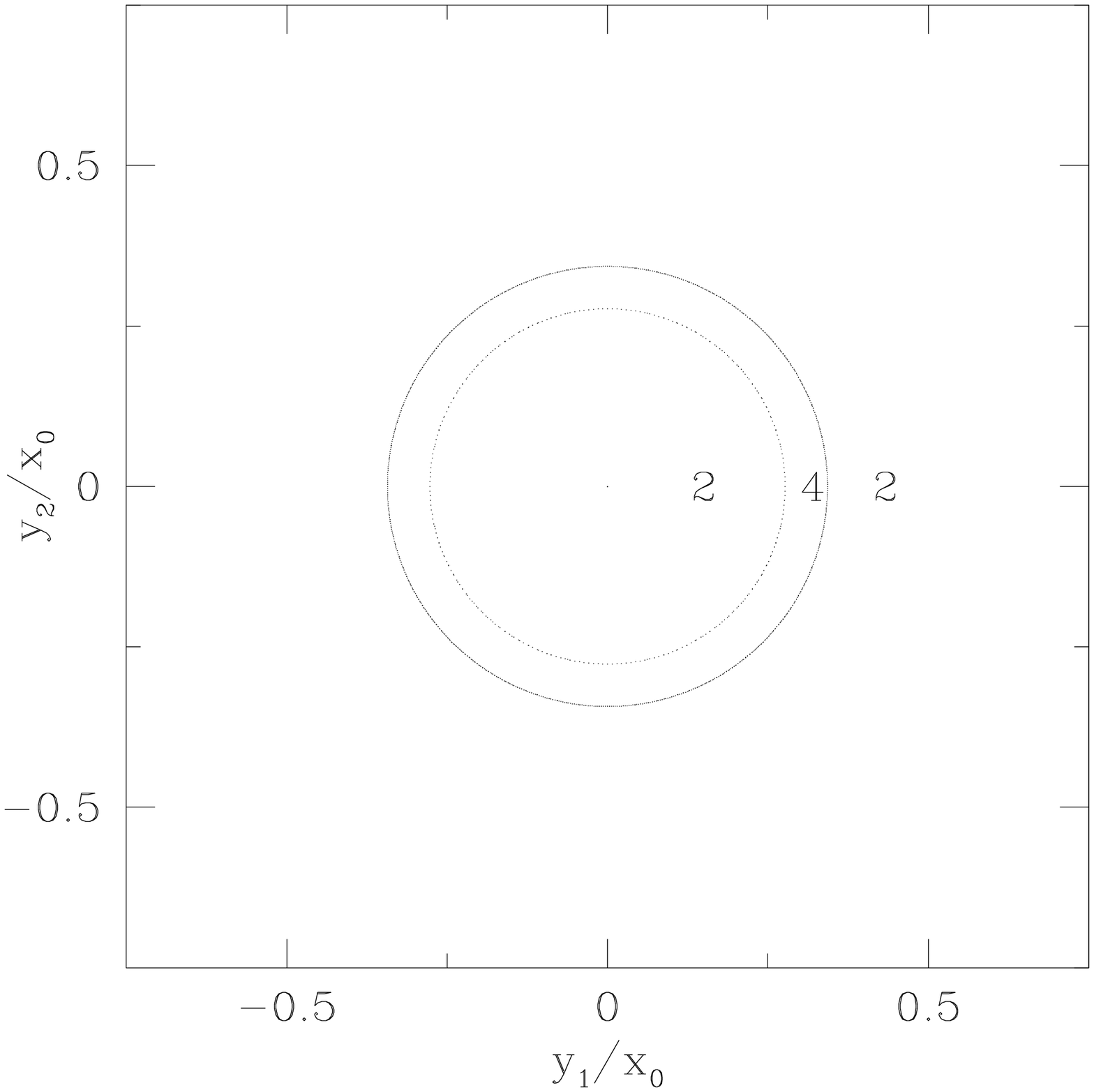}
\caption{Plummer lens with central point mass, $\kappa_0=2, \
a_0=0.02$. Left: critical circles in $L$, right: caustic circles in
$S$. Number labels indicate number of images.}
\label{fig:plum1}
\end{figure*}

\section{The Plummer Model}

Plummer's (1911) model is the simplest self-consistent solution of the
collisionless Boltzmann equation with finite total mass. Although the
density in early-type galaxies decreases less steeply than $\rho \sim
r^{-5}$ at large radii, the Plummer model still gives a good
description of the inner parts, which control the lensing properties.

For an isotropic stellar system, the distribution function $f$ depends
on the (negative) total particle energy $E$ only. The Plummer model is
then a polytrope defined by the distribution function
\begin{equation}
f(E)\propto(-E)^{7/2}.
\end{equation}
For a static system, energy conservation $E=\phi+\mathbf{v}^2/2$ holds
and, as a consequence, Poisson's equation for a self-consistent,
self-gravitating, spherical and isotropic system
\begin{equation}
\frac{1}{r^2}\frac{d}{dr}\left(r^2\frac{d\phi}{dr}\right)=16\pi^2G\int f(E) v^2dv
\end{equation}
implies the mass density (e.g. Binney \& Tremaine 1994, p. 225)
\begin{equation}
\rho=\rho_0\left(1+\left(\frac{r}{r_0}\right)^2\right)^{-5/2}
\label{plumsph}
\end{equation}
and total mass $M=4\pi\rho_0r_0^3/3$.

\subsection{Lens Equation}

The angular diameter distances from the observer to the lens plane
$L$, the source plane $S$ and from $L$ to $S$ be denoted by
$\DLO$,$\DSO$ and $\DSL$, respectively. We use the conventional
dimensionless coordinates $\mathbf{x}$ in $L$ and $\mathbf{y}$ in $S$
\citep[see e.g.,][ p. 245]{Sc99}.  The relative surface density
$\kappa_{\rm P}$ is now obtained by projecting (\ref{plumsph}) onto
$L$,
\begin{equation}
\kappa_{\rm P}=\kappa_0 \left(1+\left(\frac{x}{x_0}\right)^2 \right)^{-2}
\label{plum}
\end{equation} 
where $x=|\mathbf{x}|$ such that the origin coincides with the lens centre, $x_0=r_0/\DLO$ and
\begin{equation}
\kappa_0=\frac{M}{\pi r_0^2 \Sigma_{\rm crit}}, \qquad \Sigma_{\rm
  crit}=\frac{c^2\DSO}{4\pi G\DSL \DLO}. 
\label{kappa}
\end{equation}
The deflection potential $\Psi$ is
\begin{equation}
\Psi_{\rm P}= \Psi_0+\frac{\kappa_0x_0^2}{2}\ln\left(1+\left(\frac{x}{x_0}\right)^2\right).
\label{plumdefl}
\end{equation}
The lens equation is then
\begin{equation}
\mathbf{y}=\mathbf{x}-\nabla \Psi_{\rm P} =\mathbf{x}- \frac{\kappa_0\mathbf{x}}{1+(\frac{x}{x_0})^2}.
\label{sp3}
\end{equation}
The image and caustic properties of the Plummer lens are briefly
summarized below, for comparison with our results in \S 3 and \S 4

\subsection{Image and Caustic Properties}

The determinant of the Jacobian of the Plummer lens map is
\begin{equation}
\det J =\left(1-\frac{\kappa_0}{1\!+\!(\frac{x}{x_0})^2}\right)^2+
\left(1\!-\!\frac{\kappa_0}{1\!+\!(\frac{x}{x_0})^2}\right)\frac{2\kappa_0
(\frac{x}{x_0})^2}{(1\!+\!(\frac{x}{x_0})^2)^2}.
\label{sp5}
\end{equation}
The origins of $L$ and $S$ are critical and caustic points,
respectively. For a point source at the origin $\mathbf{y}=0$, the
lens equation implies a circular image, the Einstein ring, with radius
$x=x_0\sqrt{\kappa_0-1}$ and infinite magnification.  Otherwise,
without loss of generality, let the source position be
$\mathbf{y}=(y_1,0)$, whence $x_2=0$ and
\begin{equation}
x_1^3-y_1 x_1^2+(1-\kappa_0)x_0^2 x_1-x_0^2y_1=0.
\end{equation}
Thus, there are one or three images, depending on the source position
$y_1$, The images lie on a straight line through the lens centre. For
$y_1 \rightarrow \infty, \ J \rightarrow \ \mathrm{I}$ (the identity
matrix), so there is always one image at large $y_1$, as expected.

The critical curves can be obtained from (\ref{sp5}) with $\det J=0$
resulting in
\begin{eqnarray*}
x_{{\rm crit},1}&=&x_0\sqrt{\kappa_0-1}\\ x_{{\rm
crit},2}&=&
x_0\sqrt{\frac{\kappa_0}{2}\sqrt{1+\frac{8}{\kappa_0}}-
\left(1+\frac{\kappa_0}{2}\right)}.
\end{eqnarray*}
There are two critical circles in $L$ if $\kappa_0 > 1$. Applying the
lens equation then shows that the caustic corresponding to $x_{\rm
crit,1}$ is the origin point in $S$, whereas the other caustic is
circular, as expected by symmetry. Hence three images can occur by
crossing this caustic inwards. However, if $\kappa_0 \leq 1$ then only
a critical and caustic point exist and the number of images is one for
all $\mathbf{y} \neq \mathbf{0}$.

\subsection{Magnification Invariant}

Witt \& Mao's (1995) paper introduced the notion of a lensing
invariant. They considered a binary lens of two point masses. Their
main result is that the sum of the signed magnifications of the
maximum number of five images is unity, namely
\begin{equation}
\label{eq:wittmao}
\sum_{i=1}^5 \mu_i p_i = 1.
\end{equation}
Here, $\mu_i$ is the absolute magnification of the image, while $p_i$
is the parity. This result holds good irrespective of the position of
the source, provided it remains within the central caustic giving rise
to the maximum number of images. Subsequently, \cite{Rh97} and
\cite{Wi00} found further examples of lens models with invariants. A
general theory of such invariants using single variable complex
analysis \citep{Hu01,Ev02,An06} was developed. This approach is
used here to investigate the general circularly symmetric lens and, in
particular, the Plummer model.

\subsubsection{Circular Symmetric Lenses}

The standard complexification is introduced (e.g. Petters, Levine \&
Wambsganss 2001, chap. 15.1). Let $L_c, \ S_c$ be the complex lens and
source planes, respectively, and define the complex coordinates
\begin{equation}
z=x_1+ix_2 \in L_c, \qquad\qquad \ \zeta=y_1+iy_2 \in S_c.
\end{equation}
Let the complex conjugate be denoted by a bar. The complex lens
equation is rendered
\begin{equation}
\zeta=z-zf\left(z\bar{z} \right),
\label{leqc}
\end{equation}
where the deflection factor $f$ is
\begin{equation}
f(s)=2\frac{d\Psi(s)}{ds}.
\label{defl}
\end{equation} 
Note that, by definition, $f$ is a real-valued function so that
$\bar{f}=f$. The conjugate of the lens equation is therefore
\begin{equation}
\bar{\zeta}=\bar{z}-\bar{z}f\left( z\bar{z}\right)
\end{equation}
and so
\begin{equation}
\bar{z}=\frac{z\bar{\zeta}}{\zeta}.
\label{zbar}
\end{equation}
The geometrical meaning of (\ref{zbar}) is plain: due to circular
symmetry, we can without loss of generality orientate the real axes in
$S_c ,\ L_c$ so that the source position $\zeta=\bar{\zeta}$ is real
and hence, by (\ref{zbar}), the image positions as well: lens centre,
point source and all images are collinear. Also, (\ref{zbar}) allows
us to eliminate $\bar{z}$ from the lens equation (\ref{leqc}) so that
standard, one-dimensional complex analysis may be applied. It can be
recast thus to give the imaging equation
\begin{equation}
I\left(z_i,\zeta,\bar{\zeta}\right)=0,
\label{ieq}
\end{equation}
defined so that image positions $z_i$ correspond to the roots of
\begin{equation}
I\left(z,\zeta,\bar{\zeta}\right)=z-zf\left(\frac{z^2\bar{\zeta}}{\zeta}\right)-\zeta. 
\end{equation}
As the lens model $f$ may have isolated poles in the complex plane, the
imaging equation is meromorphic.

The signed flux magnification of the image at $z_i$ is given by
\citep[e.g.,][ p. 85]{Pe01}
\begin{equation}
\mu_i=\frac{1}{\det J(z_i)},
\end{equation}
where the Jacobian $J$ and its inverse $J^{-1}$ in complex form can be
written as \citep[e.g.,][ p. 506]{Pe01}
\[
\begin{array}{cc}
J= \left( \begin{array}{cc} \frac{\partial
\zeta}{\partial z}|_{\bar{z}} & \frac{\partial \zeta}{\partial
\bar{z}}|_{z} \\ \\ \frac{\partial \bar{\zeta}}{\partial
z}|_{\bar{z}}& \frac{\partial \bar{\zeta}}{\partial \bar{z}}|_{z}
\end{array} \right),
& J^{-1}= \left( \begin{array}{cc}
\frac{\partial z}{\partial \zeta}|_{\bar{\zeta}} & \frac{\partial
z}{\partial \bar{\zeta}}|_{\zeta} \\ \\ \frac{\partial
\bar{z}}{\partial \zeta}|_{\bar{\zeta}}& \frac{\partial
\bar{z}}{\partial \bar{\zeta}}|_{\zeta}
\end{array} \right).
\end{array}
\]
Requiring that $JJ^{-1}=\mathrm{I}$, it follows that
\begin{equation}
\det J =\frac{\frac{\partial \bar{\zeta}}{\partial \bar{z}}|_z}{ \frac{\partial z}{\partial \zeta}|_{\bar{\zeta}}}.
\end{equation}
By standard calculus, we find that the sum of the signed magnifications
of the $N$ images is
\begin{equation}
\sum_{i=1}^{N} \mu_i=\sum_{i=1}^{N} \frac{1}{\det J (z_i)}=\sum_{i=1}^{N}\frac{
-\left.\frac{\partial I}{\partial \zeta}\right|_{\bar{\zeta},z} }{
\left.\frac{\partial I}{\partial z}\right|_{\zeta,\bar{\zeta}}
\left.\frac{\partial \bar{\zeta}}{\partial \bar{z}}\right|_{z}}.
\label{mag2}
\end{equation}
By using (\ref{leqc}) in the denominator and (\ref{ieq}) in the
numerator of (\ref{mag2}), the following simplification is obtained
\begin{equation}
\sum^{N_{max}}_{i=1}\mu_i=\sum^{N_{max}}_{i=1}\frac{z_i}{\zeta
\left.\frac{\partial I}{\partial z}
\right|_{\zeta,\bar{\zeta}}}.
\label{mag3}
\end{equation}
We now seek a complex-valued function with residues of the form (\ref{mag3}). Consider the integral of
\begin{equation}
F(z, \zeta, \bar{\zeta})=\frac{z}{(\zeta+I)I},
\end{equation}
along a contour $\mathcal{C}$ enclosing all $N_{\max}$ roots $z_i$. Assuming
that the roots of the imaging equations are simple, the integrand $F$
has a simple pole at each root of $I$ and hence residues
\begin{equation}
\mathrm{res}(F; z_i)=\frac{z_i}{\zeta\left.\frac{\partial I}{\partial z}\right|_{\zeta,\bar{\zeta}}(z_i)},
\end{equation}
by Taylor expansion of $I$ about $z_i$, as required. Now from the residue theorem, we
find
\begin{equation}
\sum^{N}_{i=1}\mu_i =\sum^{N}_{i=1}\mathrm{res}(F;
z_i)=\frac{1}{2\pi i}\oint_\mathcal{C}\frac{z}{(\zeta + I)I}dz.
\label{mag4}
\end{equation}
Furthermore, two assumptions are made which both restrict the
functions $f$. First, no branch cuts to ensure singlevaluedness of $I$
are to occur, such that the complex plane is analytic everywhere
outside $\mathcal{C}$.  Secondly, it is assumed that the deflection
angle becomes negligible at large distance from the lens centre. This
amounts to $\nabla \Psi \rightarrow 0$ as $x \rightarrow
\infty$, hence $zf\rightarrow 0$ and therefore $I\rightarrow z$ as
$|z|\rightarrow \infty$. Then by Cauchy's theorem, $\mathcal{C}$ can be
distorted to a circle at infinity $\mathcal{C}_\infty$
\begin{equation}
\sum^{N_{\max}}_{i=1}\mu_i =
\frac{1}{2\pi i}\oint_\mathcal{C_\infty}\frac{z}{(\zeta + I)I}dz=\frac{1}{2\pi i}
\oint_{\mathcal{C}_\infty}\frac{dz}{z}=1.
\label{miv}
\end{equation}
This result is a magnification invariant for the maximum number of
roots $N_{\max}$ of the imaging equation.

Although each image corresponds to a complex root of the imaging
equation, there may exist roots which do not correspond to physical
images and which are therefore called spurious roots. To understand
their impact on (\ref{miv}), consider two of their properties. First,
by choosing without loss of generality $\zeta=\bar{\zeta}$ in the
imaging equation (\ref{ieq}), it is seen that both $z$ and $\bar{z}$
are solutions. So, all spurious roots occur as complex conjugate
pairs. Secondly, from (\ref{ieq}) and (\ref{mag3}), we see that the
magnifications of a pair of conjugate spurious roots are also complex
conjugate. This implies that the total contribution from spurious
roots to the signed magnification sum is always real.

\subsubsection{The Particular Case of the Plummer Model}
The Plummer lens equation (\ref{sp3}) can be rewritten as
\begin{equation}
\zeta=z-\frac{\kappa_0z}{1+z\bar{z}}.
\end{equation}
From the definition of the deflection factor $f$, it follows that
\begin{equation}
f(z)=\frac{\kappa_0}{1+\frac{z^2\bar{\zeta}}{\zeta}}.
\end{equation}
Hence $zf\rightarrow 0$ as $|z|\rightarrow \infty$ and $f$ involves no
roots of $z$ and hence no branch cut. The imaging equation is of third
degree in $z$ and, since the maximum number of images is three, there
are no spurious roots. Thus, the Plummer model has a magnification
invariant
\begin{equation}
\sum^{3}_{i=1}\mu_i p_i=1.
\end{equation}
In other words, the sum of the signed magnifications $\mu_i p_i$ of
the images is an invariant of the model. This result appears not to
have been realized before.

\section{The Plummer Model with Black Hole}
\subsection{Lens Equation}

The standard Plummer lens is now modified by introducing a point
mass $m$ as a model for a black hole at the centre. Using
dimensionless coordinates in $L$ as before, its relative surface
density is
\begin{equation}
\kappa_{\rm BH}(\mathbf{w})=\frac{m}{\DLO^2}\delta^{(2)}(\mathbf{w}),
\end{equation}
whence the total deflection potential is
\begin{eqnarray}
\Psi(\mathbf{x})& = & \Psi_{\rm P}(\mathbf{x})+\frac{1}{\pi}\int_L
\kappa_{\rm BH}(\mathbf{w})\ln|\mathbf{w}-\mathbf{x}|d^2w\noindent \\
& = & \Psi_{\rm P}(\mathbf{x})+a\ln x,
\label{potential}
\end{eqnarray}
where $a=m/\pi \DLO^2 \Sigma_{\crit}$. The lens equation follows
immediately,
\begin{equation}
\mathbf{y}=\mathbf{x}-
\frac{\kappa_0\mathbf{x}}{1+(\frac{x}{x_0})^2}-\frac{a\mathbf{x}}{x^2}
\label{pbhleq}
\end{equation}
using (\ref{plumdefl}).

\subsection{Image and Caustic Properties}

This lens equation shows that the point mass introduces an additional
image at $\mathbf{x}\rightarrow -a\mathbf{y}/y^2$ as $y\rightarrow
\infty$: the minimum number of images is two. Moreover,
eq~(\ref{pbhleq}) is of fourth degree in $x$, so a maximum of four
images is expected. Some further insight can be gained
analytically. With the substitution
\begin{equation}
\frac{x_1}{x_0}=\xi \cos \theta \ , \qquad \frac{x_2}{x_0}=\xi \sin \theta
\ , \qquad a_0=\frac{a}{x_0^2}=\frac{m}{\pi r_0^2 \Sigma_{\crit}}
\end{equation}
where $\xi \geq 0$, the Jacobian can be constructed from the partial
derivatives,
\begin{eqnarray*}
\frac{\partial y_1}{\partial
x_1}&=&1-\frac{\kappa_0}{1+\xi^2}+\frac{2\kappa_0\xi^2\cos^2\theta}
{(1+\xi^2)^2}+\frac{a_0(2\cos^2\theta-1)}{\xi^2}
\\ \frac{\partial y_1}{\partial
x_2}&=&\frac{2\kappa_0\xi^2\cos\theta\sin\theta}{(1+\xi^2)^2}+\frac{2a_0\cos\theta\sin\theta}{\xi^2}
\ \ = \ \ \frac{\partial y_2}{\partial x_1} \\ \frac{\partial
y_2}{\partial
x_2}&=&1-\frac{\kappa_0}{1+\xi^2}+\frac{2\kappa_0\xi^2\sin^2\theta}{(1+\xi^2)^2}+
\frac{a_0(2\sin^2\theta-1)}{\xi^2}.
\end{eqnarray*}
The radii of the two critical circles are given by det $J =0$, which yields: 
\begin{equation}
1-\frac{\kappa_0}{1+\xi_{\crit,1}^2}-\frac{a_0}{\xi_{\crit,1}^2}=0 \\
\end{equation}
so that
\begin{equation}
\xi_{\crit,1}=\sqrt{\frac{\kappa_0+a_0 -1}{2}\pm 
\sqrt{\left(\frac{1-(\kappa_0+a_0)}{2} \right)^2+a_0}}
\label{pbhcrit1}
\end{equation}
and
\begin{equation}
1-\frac{\kappa_0}{1+\xi_{\crit,2}^2}+\frac{2\kappa_0\xi_{\crit,2}^2}{(1+\xi_{\crit,2}^2)^2}+
\frac{a_0}{\xi_{\crit,2}^2}=0.
\label{pbhcrit2}
\end{equation}
The critical circle (\ref{pbhcrit1}) has precisely one solution for
all positive $\kappa_0,a_0$. Applying the lens equation shows that it
maps to a caustic point at the origin of $S$. If $\kappa_0 \leq 1$,
there is hence one critical circle in $L$ at $\xi_{\crit,1}$ and one
caustic point in $S$: the number of images is always two for
$\mathbf{y} \neq \mathbf{0}$. If $\kappa_0>1$, $\xi_{\crit,2}$
introduces two further critical circles in $L$ and two caustics in
$S$, in contrast to one for the Plummer model. Although
(\ref{pbhcrit2}) is a third degree polynomial in $\xi_{\crit,2}^2$ and
could in principle be solved analytically, it is, however, more instructive to
display the critical circles and caustics numerically. Figure
\ref{fig:plum1} illustrates this case with three critical circles and
two caustic circles, giving rise to an annular domain with four images
inside and two outside of it.

\subsection{Maximum Black Hole Mass}
\begin{figure*}
\centering
\includegraphics[width=0.65\columnwidth, height=0.65\columnwidth]{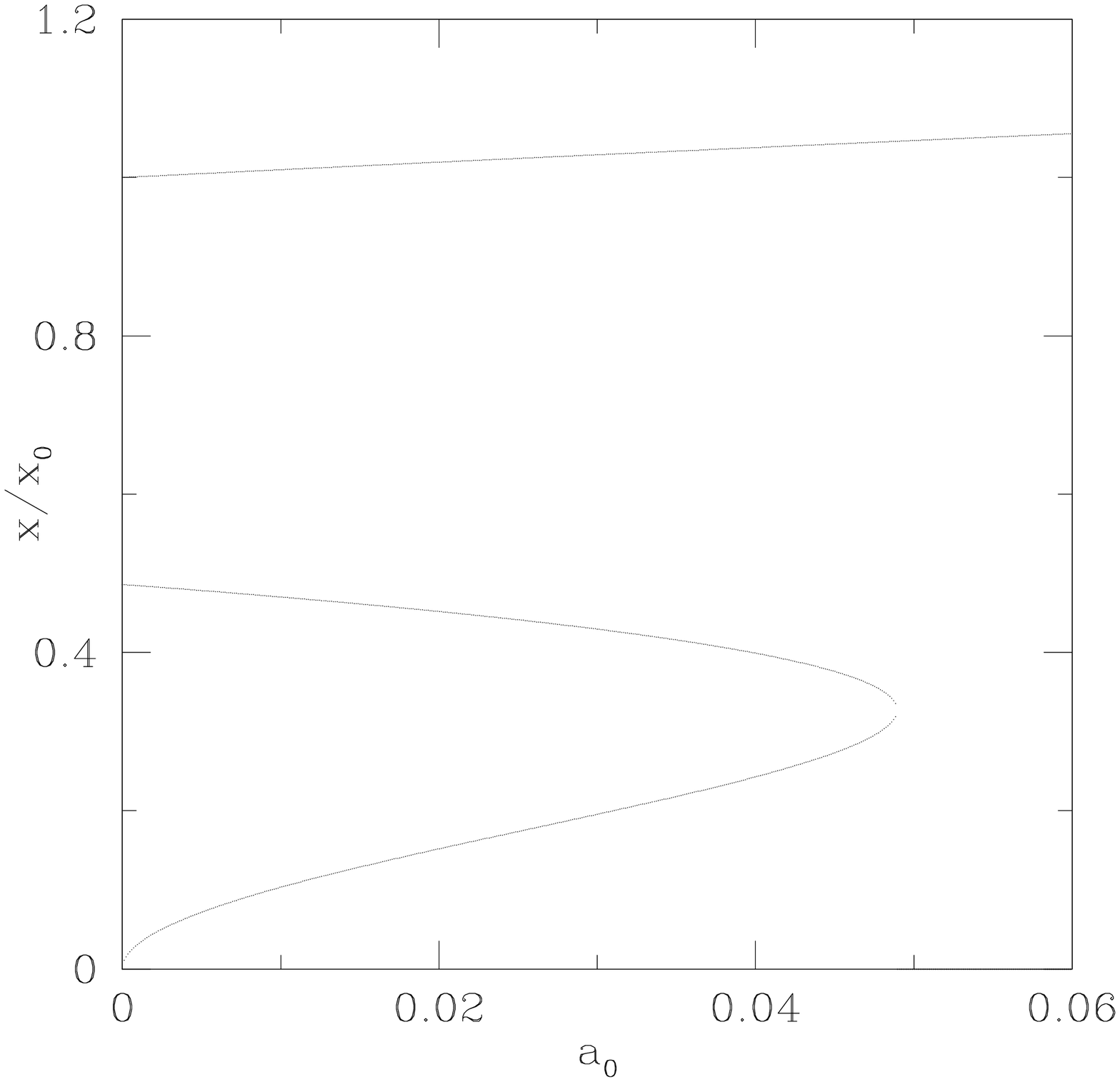}
\includegraphics[width=0.65\columnwidth, height=0.65\columnwidth]{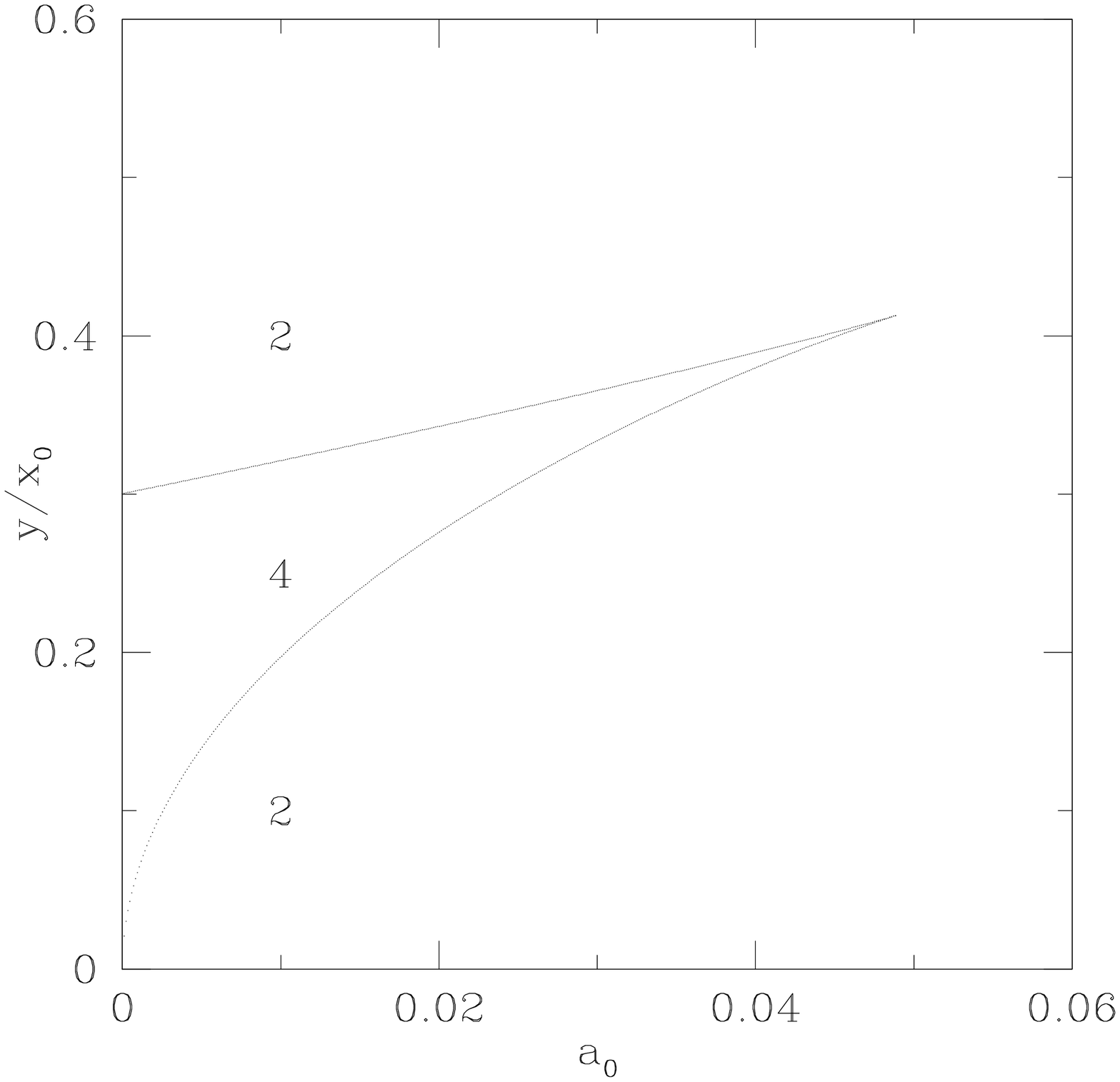}
\caption{Plummer lens with central point mass, $\kappa_0=2$. Left:
radii of critical circles in $L$ as function of point mass $a_0$,
right: radii of caustic circles in $S$ as function of $a_0$. Number
labels indicate number of images. Note the disappearance of the
caustic annulus with four images.}
\label{fig:plum2}
\end{figure*}

If the point mass $a_0$ increases while the Plummer mass $\kappa_0$ is
kept fixed, the former will eventually dominate the latter, for some
$a_0>a_{0,\max}$. In consequence, two images will occur for all
$\mathbf{y} \neq \mathbf{0}$, as is plain from the lens equation for
negligible $\kappa_0$. The annulus in $S$ resulting in four images
will hence disappear, as shown in Figure \ref{fig:plum2}.

Let us take without loss of generality $\mathbf{y}=(y_1,0) \Rightarrow
x_2=0$, then (\ref{pbhleq}) requires for the presence of four images
$\kappa_0>1$ and four intersections of
\begin{equation}
\xi_1-\frac{y_1}{x_0}=\frac{\kappa_0\xi_1}{1+\xi_1^2}+\frac{a_0}{\xi_1}.
\end{equation}
where $\xi_1=x_1/x_0$. At the threshold  $\xi_1=\xi_0,\ a_0=a_{0,\max}$, the graphs of
left-hand side and right-hand side match in
first and second derivatives,
\begin{eqnarray}
1&=&\frac{\kappa_0(1-\xi_0^2)}{(1+\xi_0^2)^2}-\frac{a_{0,\max}}{\xi_0^2}\\
0&=&\frac{2\kappa_0\xi_0(\xi_0^2-3)}{(1+\xi_0^2)^3} +\frac{2a_{0,\max}}{\xi_0^3}.
\label{deriv}
\end{eqnarray}
Eliminating $a_{0,\max}$, they yield a third degree polynomial in
$\xi_0^2$ which becomes, after the standard substitution
$q=\xi_0^2+1$,
\begin{equation}
q^3+3\kappa_0q-4\kappa_0=0.
\label{cubic}
\end{equation}
Since its cubic determinant is positive, there is one real solution
which can be obtained from Cardano's formula,
\begin{equation}
q=(2\kappa_0+\kappa_0\sqrt{\kappa_0+4})^{1/3}-(\kappa_0\sqrt{\kappa_0+4}-2\kappa_0)^{1/3}
\end{equation}
to give, using (\ref{deriv}) and (\ref{cubic}),
\begin{equation}
a_{0,\max}=\frac{(q-1)^2(4-q)}{4-3q}.
\end{equation}
In physical units, the maximum mass of the point mass for four images
relative to the Plummer model is
\begin{equation}
\frac{m_{\max}}{M}=\frac{a_{0,\max}}{\kappa_0}.
\label{maxbh}
\end{equation}
Now consider this as a model of a galaxy with a supermassive black
hole. Since many parameters are free, let us take, as a typical
example, the total galaxy mass to be $M=10^{12} \ \mathrm{M}_\odot$
and scale length $r_0=10 \ \mathrm{kpc}$. Let the lens and source be
at redshifts $z=0.5$ and $z=3$, respectively, in an Einstein-de Sitter
universe with $H=70 \ \mathrm{km \ s^{-1} \ Mpc^{-1}}$. This is a
sufficient approximation to the Concordance Cosmology for the purpose
of estimating parameters here, especially because the source distance
at higher redshift almost cancels out. A well-known expression for
angular diameter distances \citep[e.g.,][ p. 103] {Pe03}, then yields
$\DSO, \DSL, \DLO$. Hence from the definitions (\ref{kappa}), we have:
\begin{equation}
\kappa_0=\frac{4GM}{c^2r_0}\frac{\DSL\DLO}{\DSO r_0}=1.28 \ >1
\label{kappaeqn}
\end{equation}
so that four images can occur, provided the mass of the central black
hole is less than
\begin{equation}
\frac{m_{\max}}{M}=4.2\times 10^{-3} \ \Rightarrow m_{\max}=4.2\times 10^9 \ \mathrm{M}_\odot
\end{equation}
from ($\ref{maxbh}$). In fact, this result is of the same order of
magnitude as the masses inferred from observations: the presently
surmised median mass of supermassive black holes is $(1.4 \pm
0.4)\times 10^{-3}$ of the bulge mass, and galaxy M87 is
thought to have a supermassive black hole of mass $(3 \pm 1)\times 10^9 \
\mathrm{M}_\odot$ \citep{Ha04}. This indicates that by ascertaining a
fourth lensed image, a relevant upper bound on the mass of the
supermassive black hole can be obtained. Note also that this method is
viable at cosmological distances which are inaccessible to precise
mass measurements from stellar kinematics.

However, this approach has two main difficulties: first, it requires
knowledge of $\kappa_0$ and hence the mass and size of the lens, and
secondly, the fourth image may be missed or be
misinterpreted. \cite{Bo04} have recently considered the lensing
properties of a cored and a broken power-law isothermal sphere with
central point mass. Because of their diverging masses, these models
are only valid centrally and hence a black hole mass bound from
lensing, as described here, is not discussed. Based on their
numerical simulation of lensing cross-sections, the authors conclude
that observations of the fourth image, as characteristic lensing
signature of a supermassive black hole, should require the next
generation of radio telescopes but nevertheless recommend searching
already.

\subsection{Magnification Invariant}
Using the same conventions as in \S 2.3, the lens equation
(\ref{pbhleq}) can be recast thus
\begin{equation}
\zeta=z-\frac{\kappa_0z}{1+z\bar{z}}-\frac{a_0}{\bar{z}}
\end{equation}
such that
\begin{equation}
f(z)=\frac{\kappa_0}{1+\frac{z^2 \bar{\zeta}}{\zeta}}+\frac{a_0\zeta}{z^2\bar{\zeta}},
\end{equation}
so again $zf\rightarrow 0$ as $|z|\rightarrow \infty$ and $f$ has no
branch cut. The maximum number of images is four. Likewise, the
imaging equation is of fourth degree in $z$ so there are no spurious
roots. Therefore, the magnification invariant holds for the Plummer
model with central point mass as well,
\begin{equation}
\sum^{4}_{i=1}\mu_i p_i=1.
\label{invariant}
\end{equation}
Although the central black hole has yielded a further image, it is
still true that the sum of the signed magnifications is unity.

\section{The Plummer Lens with Black Hole and Shear}

\subsection{Lens Equation}

Circular symmetry is however normally broken in gravitational lensing.
This may be caused by flattening of the lensing galaxy itself or by
the presence of an external gravitational field, for instance, in a
galaxy cluster. For simplicity, let us consider a constant external
mass surface density $\kappa_{\rm ext}$ and shear $\gamma_{\rm ext}$.
Let the Cartesian coordinate axes be in the principal directions
so that the perturbation of the lens equation is in diagonal form \citep[e.g.,][ p. 250]{Sc99}
\begin{eqnarray}
\Delta\mathbf{y} & = &
-\left( \begin{array}{cc}
\kappa_{\rm ext}\!+\!\gamma_{\rm ext} & 0 \\
0 & \kappa_{\rm ext}\!-\!\gamma_{\rm ext} 
\end{array} \right) \mathbf{x} \nonumber \\
& = & 
-\left( \begin{array}{cc}
\Gamma_1 & 0 \\
0 & \Gamma_2 
\end{array} \right)\mathbf{x}
\end{eqnarray}
where $\Gamma_1,\ \Gamma_2$ are the components of the total external shear. The lens
equation (\ref{pbhleq}) is now modified thus
\begin{eqnarray}
y_1 &=& x_1-\frac{\kappa_0x_1}{1+(\frac{x}{x_0})^2}-\frac{ax_1}{x^2}-\Gamma_1x_1 
\label{shear1}
\\
y_2 &=& x_2-\frac{\kappa_0x_2}{1+(\frac{x}{x_0})^2}-\frac{ax_2}{x^2}-\Gamma_2x_2.
\label{shear2}
\end{eqnarray}
For $\Gamma_1 \neq \Gamma_2$, the imaging equation is a fifth order
polynomial in $z \bar{z}$. However, as we will see in the next section, the maximum
number of images of a Plummer model with black hole and shear is 6.
Hence in general, there are spurious roots to the imaging polynomial and a
magnification invariant does not exist.

\subsection{Image and Caustic Properties}
\begin{figure}
\centering \includegraphics[width=0.45\columnwidth,
height=0.45\columnwidth]{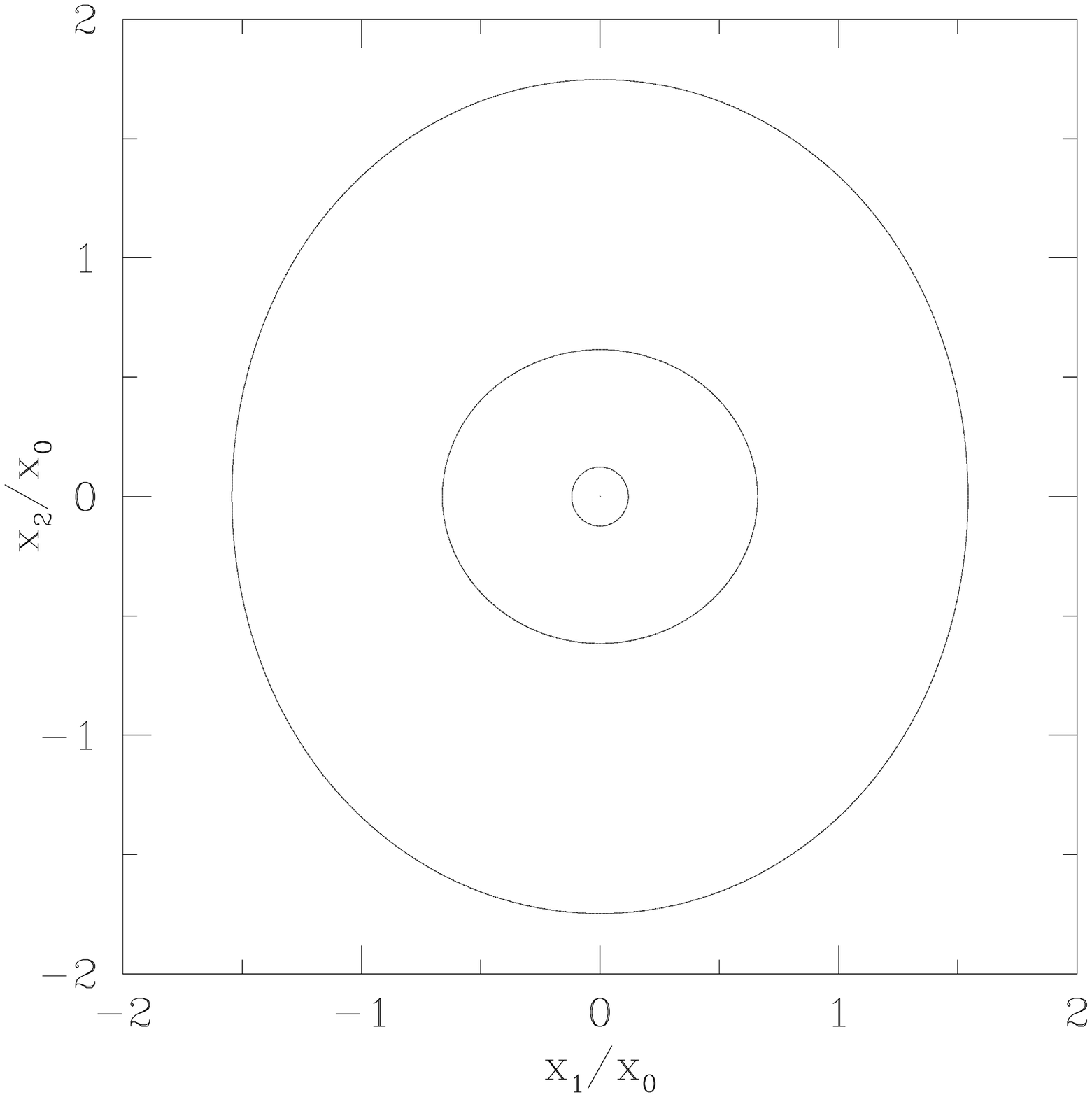}
\includegraphics[width=0.45\columnwidth,
height=0.45\columnwidth]{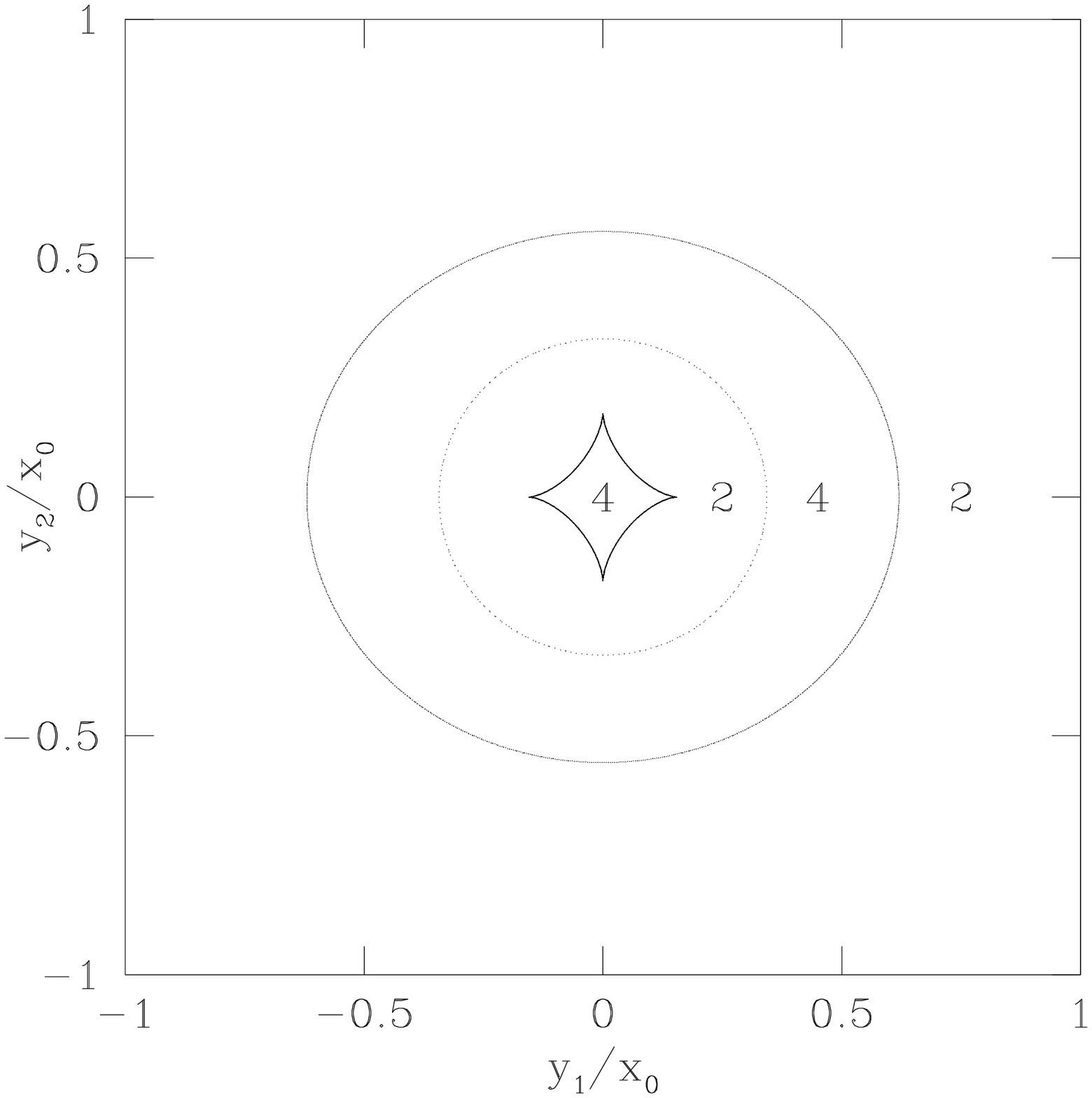}
\includegraphics[width=0.45\columnwidth,
height=0.45\columnwidth]{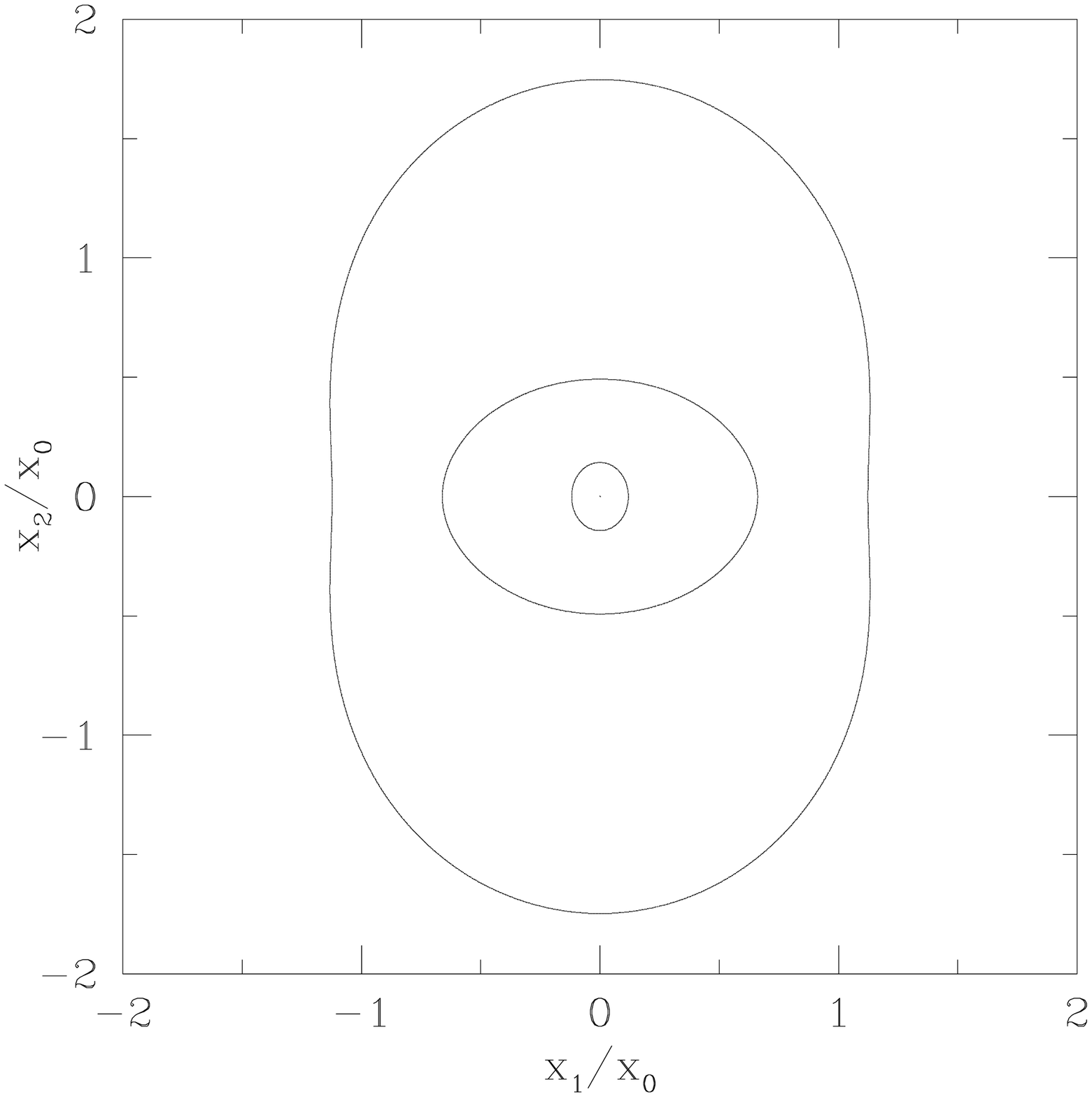}
\includegraphics[width=0.45\columnwidth,
height=0.45\columnwidth]{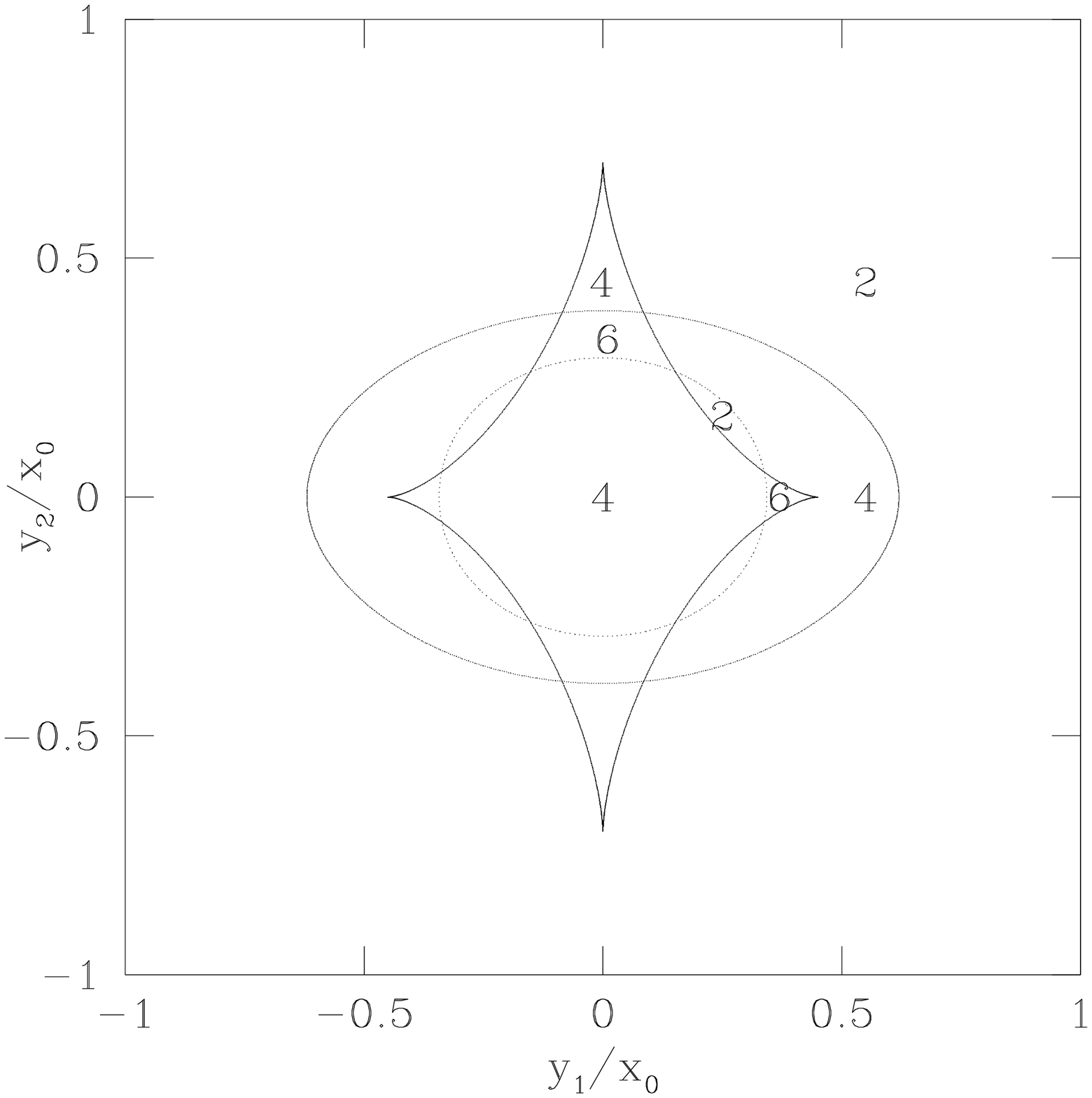}
\includegraphics[width=0.45\columnwidth,
height=0.45\columnwidth]{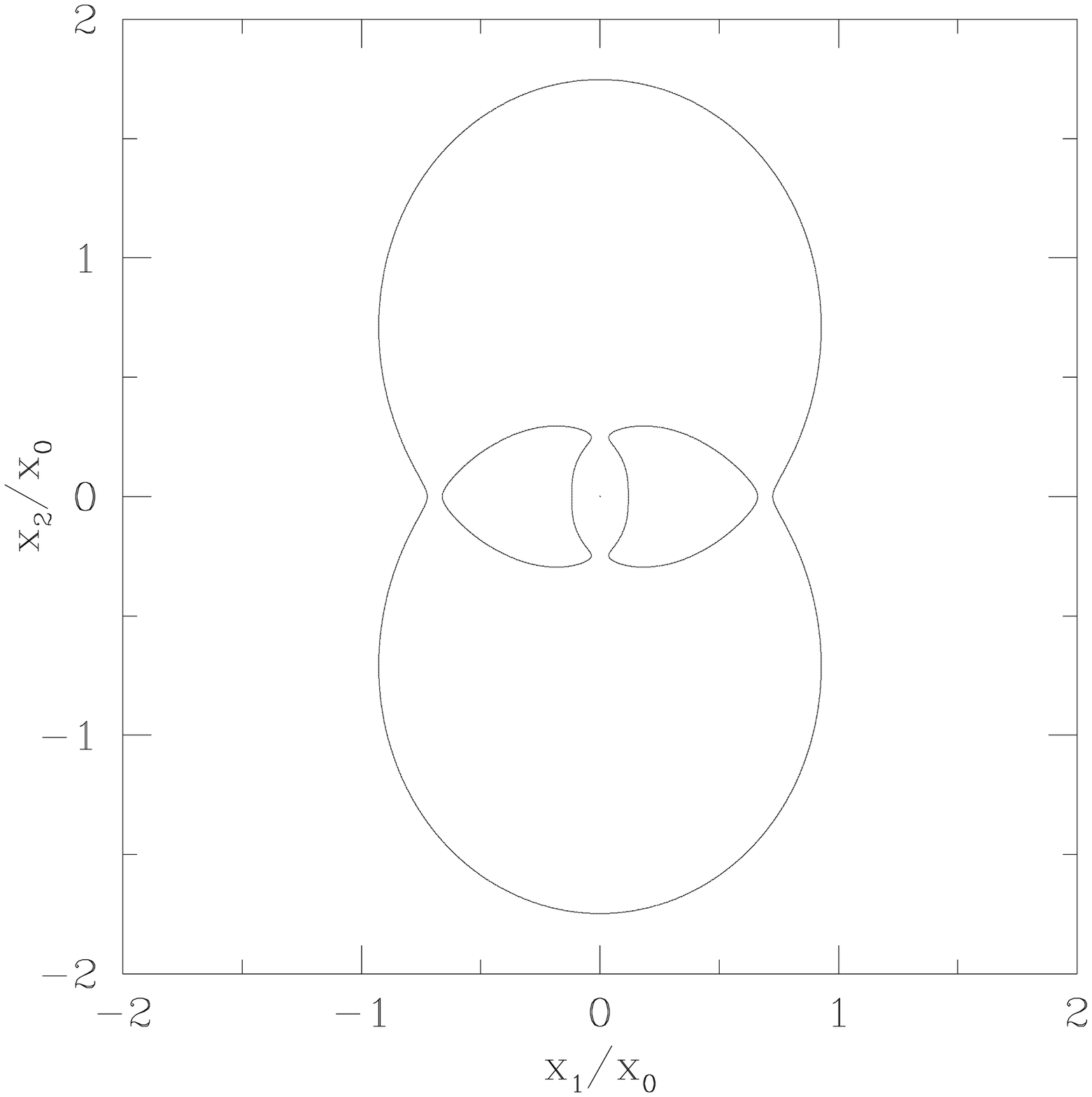}
\includegraphics[width=0.45\columnwidth,
height=0.45\columnwidth]{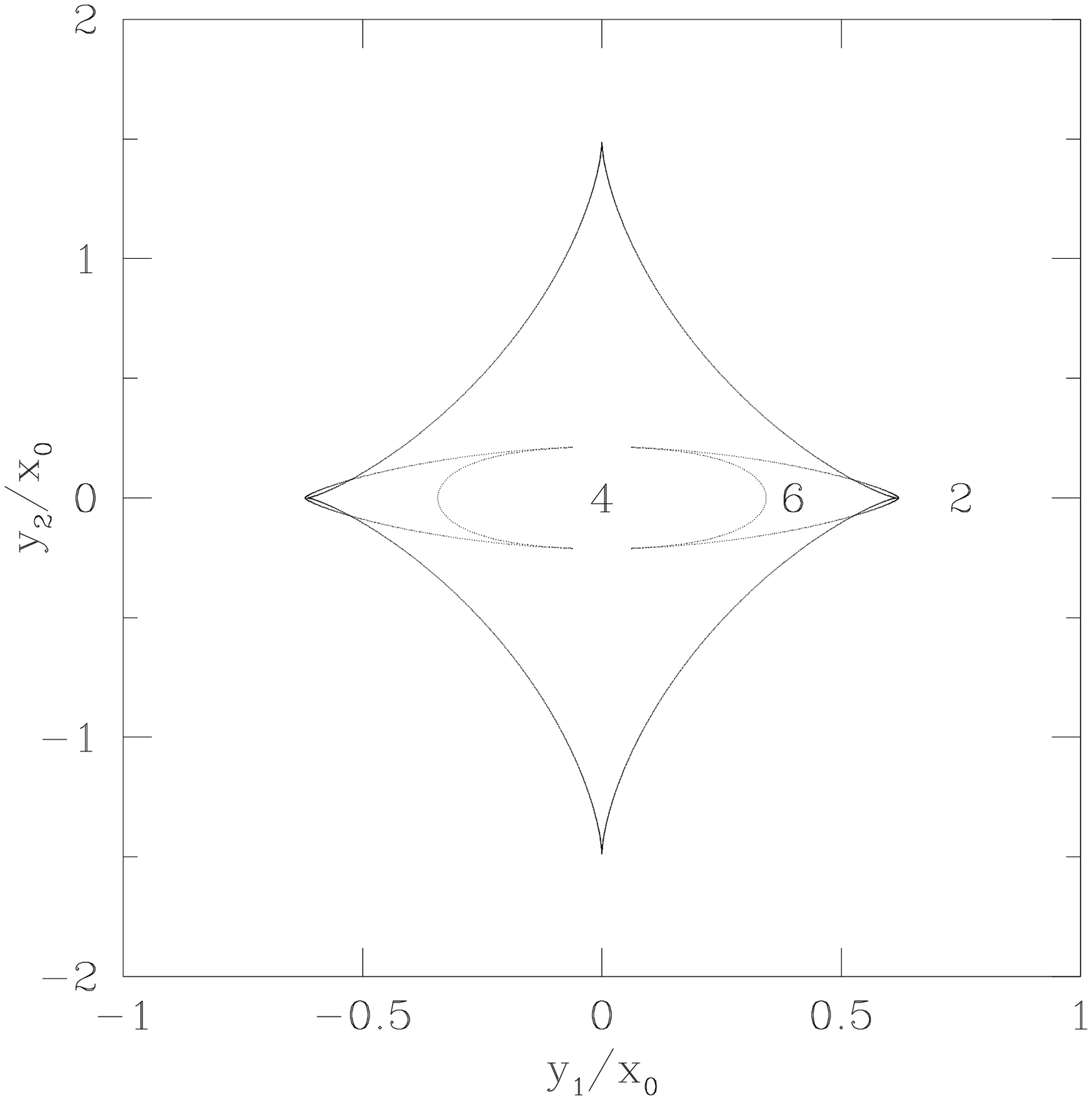}
\caption{Plummer lens with point mass and external shear, $\kappa_0=2,
\ a_0=0.02$. The difference between the components of external shear
increases from top to bottom. Upper left: critical curves, upper
right: caustics, for $\Gamma_1=0.5, \ \Gamma_2=0.4$. Middle left:
critical curves, middle right: caustics, for $\Gamma_1=0.5, \
\Gamma_2=0.1$. Lower left: critical curves, lower right: caustics, for
$\Gamma_1=0.5, \ \Gamma_2=-0.35$. Number labels indicate number of
images.}
\label{fig:plum3}
\end{figure}
\subsubsection{Astroid Caustic}

When $\Gamma_1=\Gamma_2$, eqns (\ref{shear1}) and (\ref{shear2})
remain circularly symmetric with three critical circles, two caustic
circles and a critical and caustic point, as in \S 3.

Consider now the case $\Gamma_1 \neq \Gamma_2, \ \Gamma_1>0 \wedge
\Gamma_2>0$, that is, the external mass surface density $\kappa_{\rm
ext}>\gamma_{\rm ext}$. Since, by construction, the coordinate axes
point in the principal directions of the external shear, the critical
curves in $L$ intersect the $x_1$-axis with a tangent vector parallel
to $(0,1)$ and the $x_2$-axis with $(1,0)$. The corresponding caustic
tangent vanishes if, in addition, $\partial y_2/\partial x_2=0$ or
$\partial y_1/\partial x_1=0$, respectively. Then, these are
\textit{cusp points} on the $y_1,y_2$-axes in $S$
(e.g. \citep{Sc99}, p. 252). They may be categorised thus:
\begin{itemize}
\item{Cusp points on $y_1$-axis: $x_2=0 \wedge \frac{\partial
y_2}{\partial x_2}=0$
\begin{eqnarray*}
\Rightarrow &0=&
1-\frac{\kappa_0}{1-\Gamma_2}\frac{1}{1+(\frac{x_1}{x_0})^2}-\frac{a}{1-\Gamma_2}\frac{1}{x_1^2}
\ \ \mathrm{from \ \ (\ref{shear2})} \\ \Rightarrow
&y_1=&x_1(\Gamma_2-\Gamma_1) \ \ \mathrm{from \ \ (\ref{shear1})}
\end{eqnarray*}
There is precisely one solution for $x_1^2$, hence precisely two for
$x_1$ and two for $y_1$ since $\Gamma_1 \neq \Gamma_2$.  }
\item{Cusp points on $y_2$-axis: $x_1=0 \wedge \frac{\partial y_1}{\partial x_1}=0$
\begin{eqnarray*}
\Rightarrow &0=&
1-\frac{\kappa_0}{1-\Gamma_1}\frac{1}{1+(\frac{x_2}{x_0})^2}-\frac{a}{1-\Gamma_1}\frac{1}{x_2^2}
\ \ \mathrm{from \ \ (\ref{shear1})} \\ \Rightarrow
&y_2=&x_2(\Gamma_1-\Gamma_2) \ \ \mathrm{from \ \ (\ref{shear2})}
\end{eqnarray*}
Again, there is precisely one solution for $x_2^2$, hence precisely two
for $x_2$ and two for $y_2$ since $\Gamma_1 \neq \Gamma_2$.}
\end{itemize}

Therefore, $\Gamma_1 \neq \Gamma_2$ lifts the degeneracy of the
caustic point at $\mathbf{y}=\mathbf{0}$ which was found the previous
circularly symmetric cases. There is now one pair of cusp points on
each of the $y_1,y_2$-axes giving rise to an \textit{astroid}
caustic. By crossing the astroid inwards, the number of images
increases from two to four. If the point source is at
$\mathbf{y}=\mathbf{0}$, an additional fifth image at the centre of
$L$ is produced, since by (\ref{shear1}) and (\ref{shear2})
$\mathbf{x}=\mathbf{0}$ is a solution. Note that, in this case,
circular symmetry is broken and an Einstein ring image cannot
form. The astroid increases in size as $\Gamma_1,\Gamma_2$ become more
different, eventually intersecting the already existing caustic
curves. Caustic domains are thereby created in which the image
multiplicity increases from a maximum of four to six. Such a
configuration of critical and caustic curves is illustrated in the
upper and middle panel of figure \ref{fig:plum3}.
\subsubsection{Metamorphoses}
\begin{figure*}
\centering \includegraphics[width=0.65\columnwidth,
height=0.65\columnwidth]{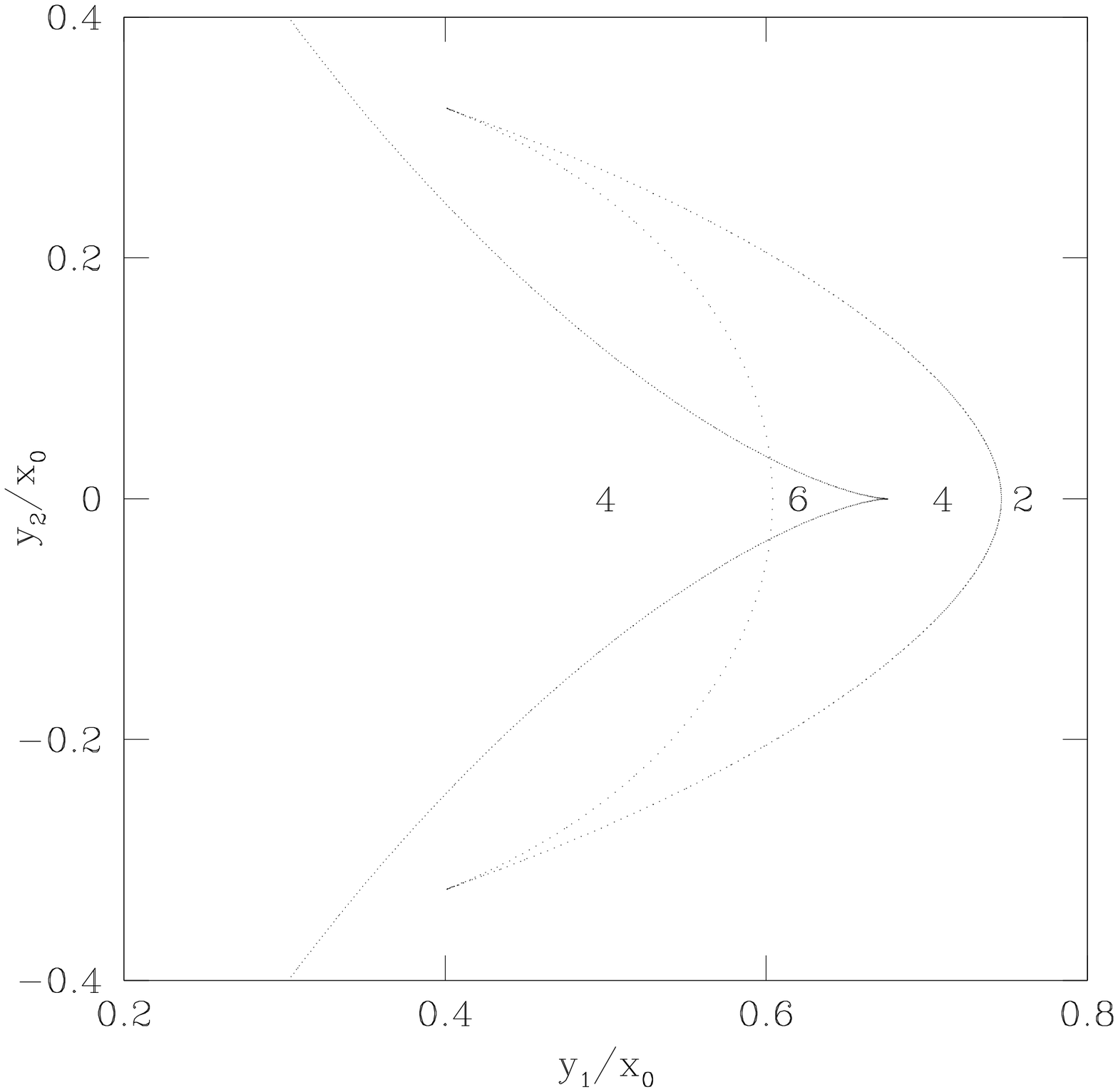}
\includegraphics[width=0.65\columnwidth,
height=0.65\columnwidth]{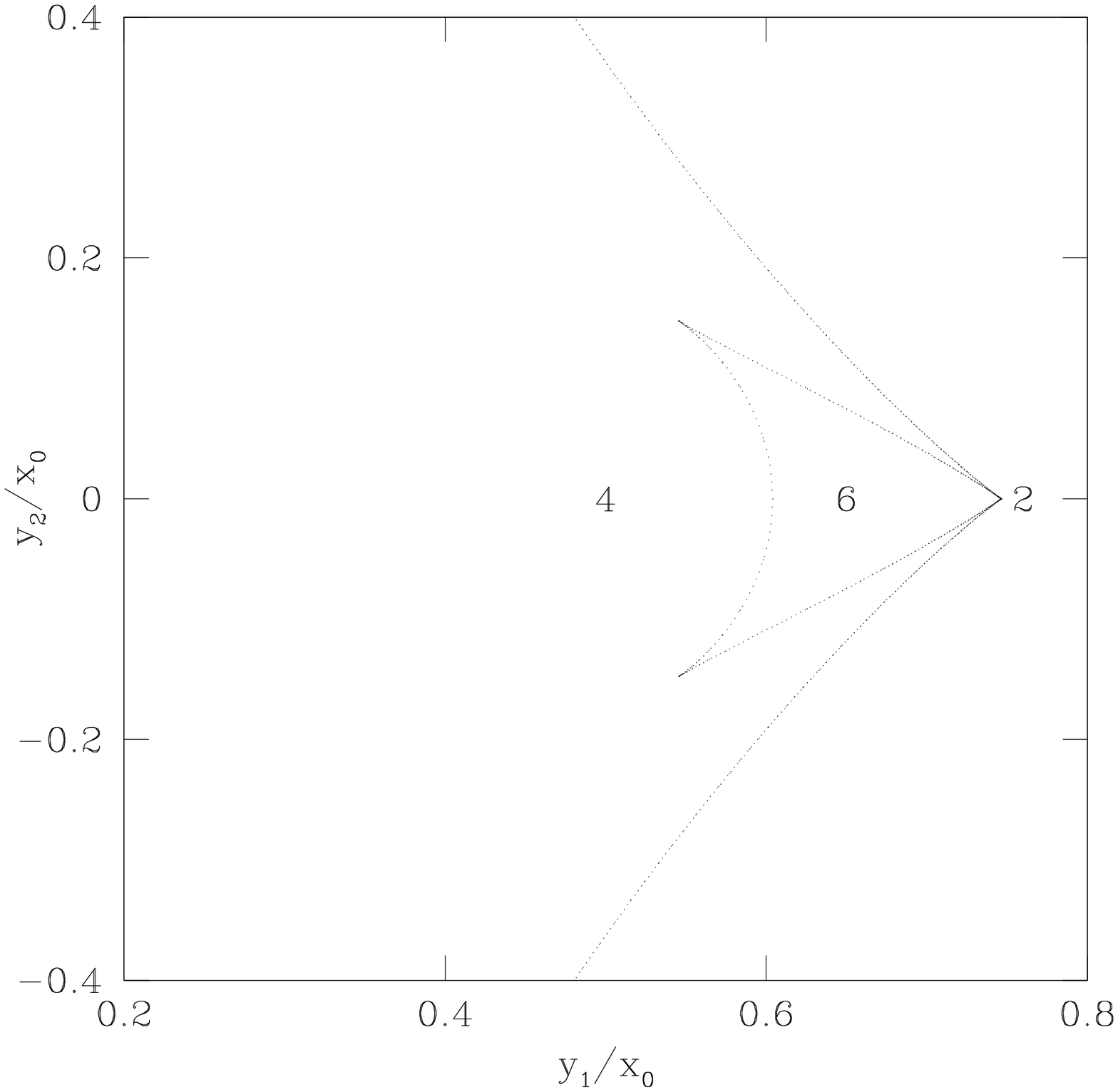}
\includegraphics[width=0.65\columnwidth,
height=0.65\columnwidth]{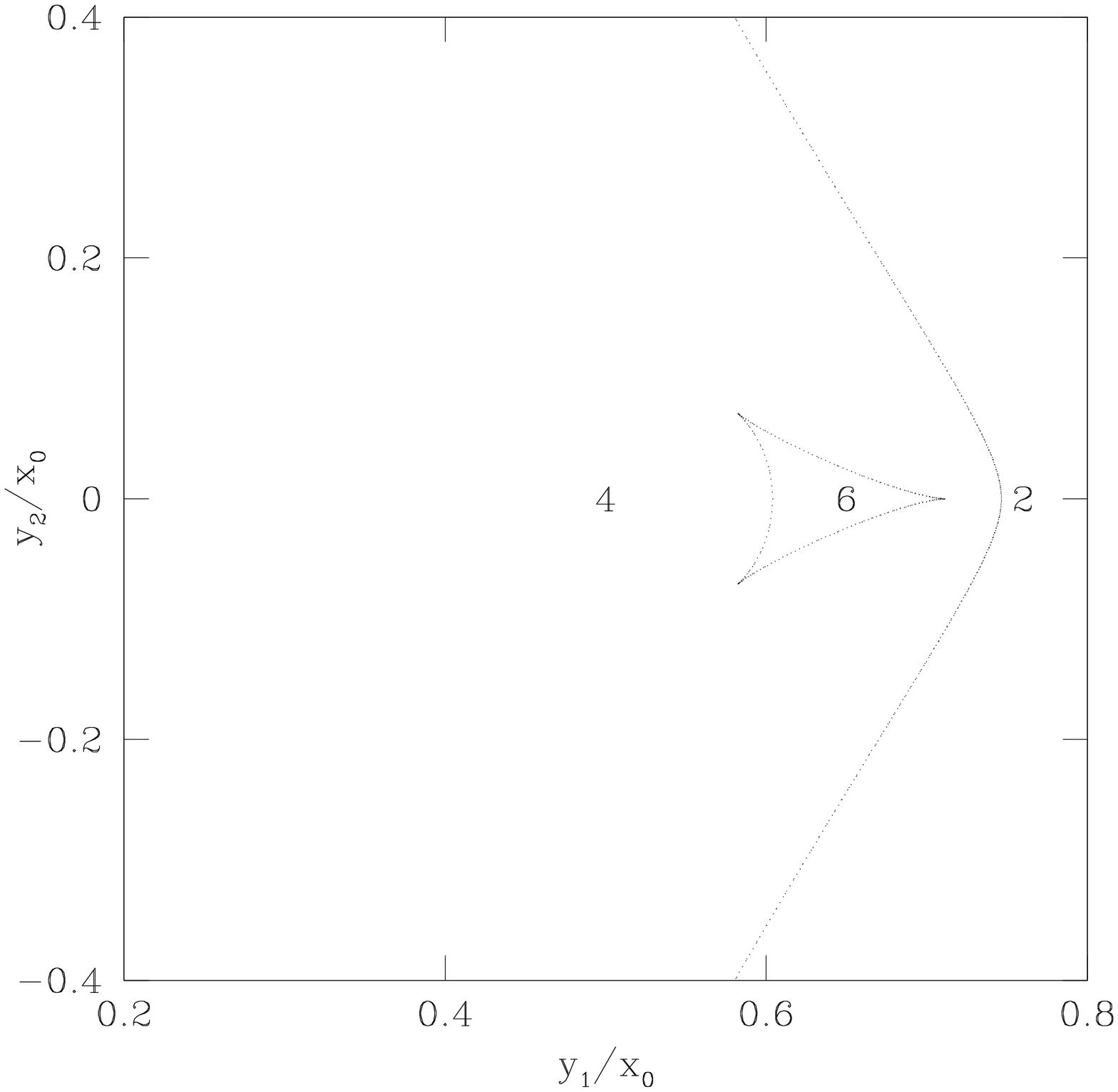}
\caption{Caustics for the Plummer lens with central point mass and
external shear, $\kappa_0=2, \ a_0=0.06, \ \Gamma_1=0.6$. The
difference between the components of external shear increases left to
right as $\Gamma_2$ decreases. Left: $\Gamma_2=-0.1$, middle:
$\Gamma_2=-0.52$ where the umbilic metamorphosis occurs, right:
$\Gamma_2=-0.9$. Number labels indicate number of images.}
\label{fig:umbilic}
\end{figure*}
As the difference in the shear components increases yet further,
$\Gamma_1>0 \wedge \Gamma_2<0$ and the external mass surface density
$\kappa_{\rm ext}<\gamma_{\rm ext}$. Now critical curves and caustics rejoin,
as can be seen in the lower panel of figure \ref{fig:plum3}: as the
inner two critical curves in $L$ merge on the $x_2$-axis, the tangent vector 
vanishes and $\nabla \det J=0$. This corresponds to the rejoining of the inner two caustics on the
$y_2$-axis to give two cusps, and is an example of a beak-to-beak
metamorphosis.  \\ Our model can also serve to illustrate the
metamorphosis of an umbilic. Referring to the deflection potential
$\Psi$ in equation (\ref{potential}) and the lens equation with
external shear (\ref{shear1}), (\ref{shear2}), the conventional form
\citep[e.g.,][ p. 162]{Sc99} of the Jacobian $J$ of the lens mapping is,
\begin{equation}
J=\left( \begin{array}{cc}
1-\kappa-\gamma_1 & -\gamma_2 \\
-\gamma_2 & 1-\kappa+\gamma_1
\end{array} \right)
\end{equation}
where 
\begin{eqnarray*}
\gamma_1&=& \frac{1}{2}\left(\frac{\partial^2 \Psi}{\partial
  x_1^2}-\frac{\partial^2 \Psi}{\partial x_2^2}\right)+\gamma_{\rm ext} \\
\gamma_2 &=& \frac{\partial^2 \Psi}{\partial x_1 \partial x_2}\\
\kappa &=& \frac{1}{2}\left(\frac{\partial^2 \Psi}{\partial
  x_1^2}+\frac{\partial^2 \Psi}{\partial x_2^2}\right)+\kappa_{\rm ext}.
\end{eqnarray*}
The umbilic point occurs where both eigenvalues $\lambda_1,\lambda_2$
of $J$ vanish,
\begin{eqnarray*}
\lambda_1 &=& 1-\kappa+\sqrt{\gamma_1+\gamma_{\rm ext}} = 0 \\
\lambda_2 &=& 1-\kappa-\sqrt{\gamma_1+\gamma_{\rm ext}} = 0.
\end{eqnarray*}
These two conditions can now be applied to derive the location of
umbilic metamorphoses. Since, by construction, the coordinate system
in the lens plane is aligned with the principal directions of
external shear, we can set $\Gamma_1 > \Gamma_2$ and $x_2=0$. Then
keeping $\kappa_0, \ a_0, \ \Gamma_1$ fixed defines a one-parameter
family of lens models in $\Gamma_2$. Letting $\xi_1=x_1/x_0$, we hence
obtain the value of $\Gamma_2$ for the metamorphosis from
\begin{eqnarray*}
\frac{\kappa_0 (1-\xi_1^2)}{(1+\xi_1^2)^2} - \frac{a_0}{\xi_1^2} &=& 1-\Gamma_1 \\
\frac{\kappa_0}{1+\xi_1^2} + \frac{a_0}{\xi_1^2} &=& 1-\Gamma_2. \\
\end{eqnarray*}
Figure \ref{fig:umbilic} illustrates the caustics for a particular
Plummer model with central point mass and external shear undergoing
metamorphosis at an umbilic point.

\section{Conclusion}

This paper provides an analytic study of the lensing properties of the
Plummer model, together with modifications incorporating the effects
of a central point mass and external shear. This is a simple model for
lensing by a flattened galaxy with a central black hole.  The maximum
number of images given by the circular Plummer lens is 3. This number
rises to 4 if a central point mass is introduced and to 6 if external
shear is added as well.

There are three new results of our work.  First, a magnification
invariant is found to exist for both the Plummer lens alone, and for
the Plummer lens with black hole. Even though the introduction of a
black hole changes the maximum number of images from 3 to 4, it is
nonetheless true that the sum of the signed magnifications remains
unity.

The second result is a treatment of the caustic structure for the
Plummer lens with black hole and shear.  Without shear, there are two
caustic circles and a caustic point. This gives three disjoint regions
in the source plane where double, quadruple and double imaging
respectively occur.  The introduction of shear causes the caustic
point to unfold as an astroid caustic.  The presence of four cusps on
an astroid caustic is established, demonstrating the change of image
multiplicities as the caustics intersect and metamorphoses occur. For
modest shear, the astroid is contained within the two outer
caustics. However, as the components of shear $\Gamma_1$ and
$\Gamma_2$ become more different, the size of the astroid increases
and additional caustic domains are created in which the image
multiplicity is 6.

Based on this discussion of caustic curves, we describe a possible 
method to obtain an upper limit on the mass of a black hole from
gravitational lensing. As this approach relies on counting images
alone, it could be used at cosmological distance and therefore as an
independent check on mass estimates obtained from other methods, such
as stellar kinematics and reverberation mapping. This emphasises the
value of lensing surveys for constraining the mass distribution in the
inner parts of galaxies.

\section*{Acknowledgments}
MCW gratefully acknowledges PPARC funding.

\label{lastpage}

\end{document}